\documentclass[twocolumn]{article}

\usepackage[utf8]{inputenc}
\usepackage{graphicx}
\usepackage{amsmath}
\usepackage{amssymb}
\usepackage{amsfonts}
\usepackage{booktabs}
\usepackage{tabularx}
\usepackage{siunitx}
\usepackage{xcolor}
\usepackage[english]{babel}
\usepackage{microtype}
\usepackage{floatrow}
\usepackage{xurl}
\usepackage[a4paper,margin=2cm]{geometry}
\usepackage{newtxtext}
\usepackage{newtxmath}

\usepackage[numbers,square,sort&compress]{natbib}
\usepackage[colorlinks=true,linkcolor=blue,citecolor=blue]{hyperref}
\hypersetup{breaklinks=true}
\usepackage{doi}

\usepackage{authblk}

\usepackage{fancyhdr}
\fancypagestyle{mystyle}{%
  \fancyhf{}%
  \fancyfoot[L]{This is the author's accepted manuscript. \\Accepted for publication in Nuclear Instruments and Methods in Physics Research Section A}%
}
\fancypagestyle{plain}{\fancyhf{}\fancyfoot[L]{This is the author's accepted manuscript. \\Accepted for publication in Nuclear Instruments and Methods in Physics Research Section A}}

\definecolor{myblue}{HTML}{1f77b4}
\definecolor{mygreen}{HTML}{2ca02c}
\definecolor{myred}{HTML}{d62728}
\definecolor{myorange}{HTML}{ff7f0e}

\title{The STRASSE liquid hydrogen target system}

\newcommand{\equalcontrib}{\textsuperscript{\ensuremath{\dagger}}}

\author[1]{M.~Enciu\equalcontrib\thanks{Corresponding author.}}
\author[1]{C.~Xanthopoulou\equalcontrib}
\author[1]{A.~Enciu}
\author[1]{H.N.~Liu}
\author[1]{A.~Obertelli}
\author[1]{A.I.~Stefanescu}
\author[1]{F.~Wienholtz}

\affil[1]{Institute of Nuclear Physics, Technische Universit\"at Darmstadt, 64289 Darmstadt, Germany}

\date{}

\begin{document}
\maketitle
\thispagestyle{fancy}
\begingroup
  \renewcommand{\thefootnote}{\ensuremath{\dagger}}\footnotetext{These authors contributed equally to this work.}
\endgroup

\begin{abstract}
A compact liquid hydrogen target system has been developed for the Silicon Tracker for RAdioactive nuclei Studies at SAMURAI Experiments (STRASSE) at the RIKEN Nishina Center. This target, designed for proton-induced quasi-free scattering measurements in inverse kinematics, features a customizable cylindrical cell with a volume up to 125~mL which increases the reaction rate/luminosity, and thin Mylar walls to minimize the protons' angular straggling. The cryogenic system, operated at 20~K, is optimized for rapid cool-down ($\leq$ 5~h) and empty-target measurements, avoiding long experimental dead time. This new setup will allow for high-precision studies of nuclear structure using both missing-mass and in-flight prompt $\gamma$-ray spectroscopy techniques at the RIBF facility.
\end{abstract}


\section{Introduction}\label{sec1}

Nuclear structure studies using radioactive ion beams have significantly advanced over the past two decades. However, accessing the very exotic neutron-rich nuclei remains experimentally challenging due to low production rates and small reaction cross-sections. To overcome these limitations, experimental setups are specifically designed to maximize the luminosity and at the same time improve the particle detection efficiency and energy resolution. In this context, in-beam $\gamma$-ray spectroscopy in combination with quasi-free scattering (QFS) reactions has proven to be a powerful approach for probing single-particle configurations and correlations in exotic nuclei~\cite{QFSreview}.\par
A major contribution to the latest advances in the studies of exotic nuclei is the use of liquid hydrogen (LH$_2$) targets. LH$_2$ targets offer a clean reaction environment with minimal background from heavier elements, while having a density of 0.071~$g/cm^{3}$ (1~atm, 20~K), about 850 times denser than hydrogen gas at room temperature and atmospheric pressure. It allows precise background subtraction using empty-target measurements during the experiment, without long beam-time interruption. The LH$_2$ targets can be manufactured in a wide range of lengths limited only by the available cooling power, enabling fine-tuning of the target geometry to meet the phase-space requirements of each physics case.

Several LH$_2$ target systems have been developed for the large radioactive ion beam facilities, tailored to different detection geometries and nuclear physics goals~\cite{AO_reviewlh2}. For instance, the COCOTIER system~\cite{COCOTIERref} at R3B/GSI was designed for QFS measurements in inverse kinematics and accommodates various target lengths up to 150 mm within a fixed 42 mm diameter. Another system, developed within the PRESPEC project~\cite{PRESPECpaper}, was optimized for in-beam $\gamma$-ray spectroscopy with arrays like AGATA at GSI/FAIR. At RIKEN, the MINOS device~\cite{MINOSpaper} was developed to support the in-beam $\gamma$-ray spectroscopy program using the SAMURAI and ZeroDegree spectrometers. It integrates a thick LH$_2$ target (of 150 mm length and 52 mm diameter) with a surrounding TPC for proton tracking in (p,2p) and (p,pn) reactions. \par
Based on the results achieved with the above mentioned systems, the STRASSE project~\cite{STRASSEpaper} developed a new LH$_2$-based detection setup optimized for QFS experiments at intermediate energies. STRASSE incorporates a compact and modular LH$_2$ target, surrounded by a silicon charged-particle tracker with full azimuthal angular coverage and high rate readout ($\geq$~500~kHz) capability. The target cell, with a reduced diameter of $\approx$~30~mm and Mylar walls (of $\approx$~175~$\mu$m), minimizes multiple scattering and angular straggling of the protons, improving the energy and momentum reconstruction resolution. With a target length of up to 150 mm, STRASSE achieves high luminosity while maintaining sub-millimeter vertex resolution and a missing-mass resolution below 2 MeV.\par
The design of the STRASSE LH$_2$ target system is well-suited for a broad physics program including prompt $\gamma$-ray spectroscopy, QFS-based missing-mass experiments, and the study of nucleon knockout reactions from exotic neutron-rich nuclei near the drip line. The STRASSE setup is foreseen to be used in the target area (focal plane F13) of the SAMURAI spectrometer, see Fig. 26 of Ref.~\cite{STRASSEpaper} for a detailed view of the integration of the STRASSE setup on the SAMURAI beamline at RIBF. The STRASSE system is now an integral part of an approved experimental program at RIBF, aiming to explore neutron single-particle structure, halo configurations, and deeply-bound nucleon dynamics. \par
In the following, in Section \ref{sec2} we present an overview of the cryogenic target system. Section \ref{sec3} highlights the parameters optimization and the operation conditions. For comparison, a thermosiphon model was created and it is presented in Section \ref{sec4}. Finally, the results of in-beam commissioning experiment of the LH$_2$ system coupled with the TOGAXSI system~\cite{Tanaka2023} are listed in Section \ref{sec5}. \par
\begin{figure}[ht]
    \centering
    \includegraphics[width=1\linewidth]{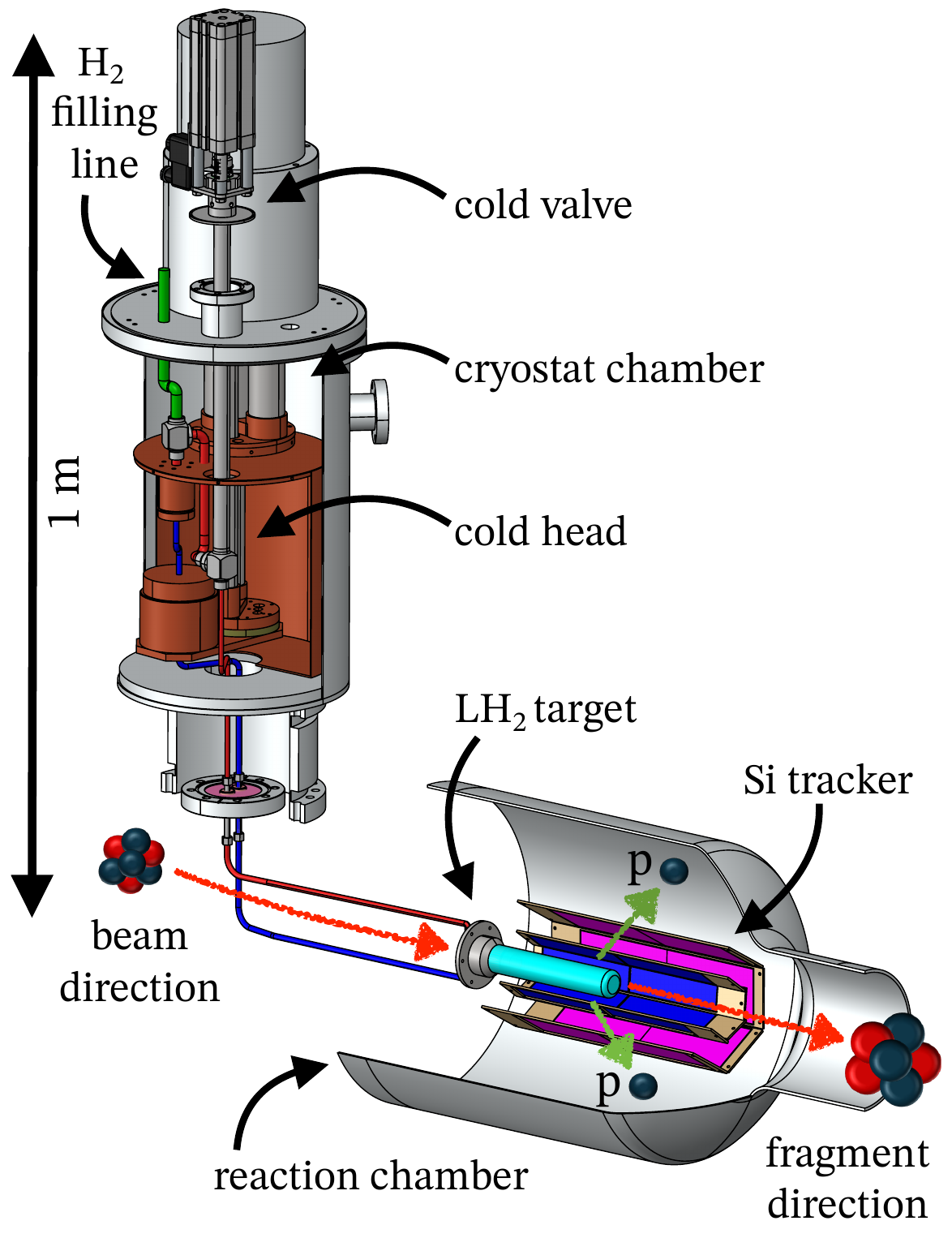}
    \caption{Drawing of the STRASSE liquid hydrogen target surrounded by the STRASSE silicon tracking system. The target cell is the central component where the proton-induced reactions take place. The cryostat is placed upstream from the reaction chamber, above the beam line. Beamline components were removed for better visibility and a cross section of the cryostat and the reaction chamber is displayed. The beam and fragment direction are marked on the plot, as well as the trajectories of the protons after a (p,2p) reaction. The cryogenic components are marked on the figure and are later discussed in the text.}
    \label{fig:STRASSE}
\end{figure}
\begin{figure*}[t]
    \centering
    \includegraphics[width=0.7\linewidth]{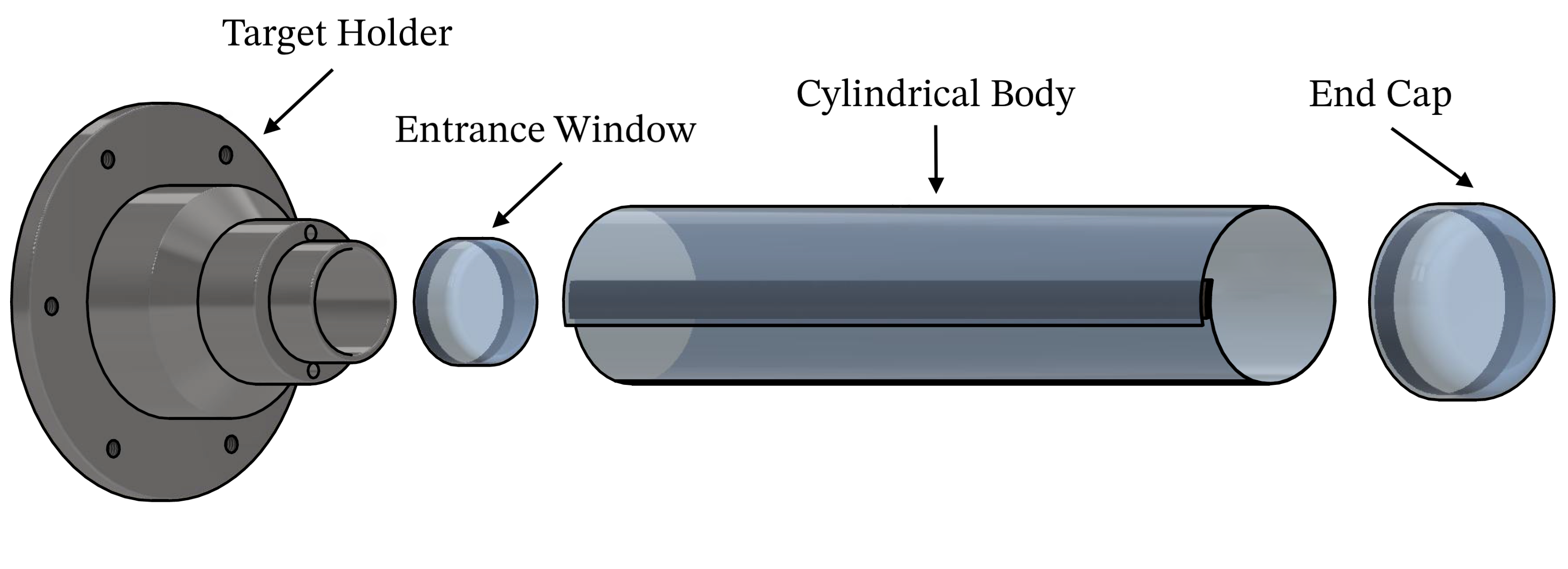}
    \caption{Drawing of the target holder and the target cell components. The target cell is composed of an entrance window, the cylindrical body and the end cap. The three components are glued together and on the target holder as indicated by the gray bands.  }
    \label{fig:target-cell}
\end{figure*}
\section{Technical design of the liquid hydrogen target}\label{sec2}
The liquid hydrogen target system is composed of a cryostat, the target cell, and the gas handling system. Each component will be described in this section. A drawing of the cryostat and the target cell is shown in Fig.~\ref{fig:STRASSE}.\par

\subsection{Target cell}\label{subsec2-1}
The central component of the liquid hydrogen target system is the target cell which is designed to have thin walls allowing for minimum energy loss for the beam, fragments and protons exiting the liquid hydrogen target at any angle~\cite{STRASSEpaper}. The target cell presented in this paper has a cylindrical shape of customizable length (up to 150~mm effective length) and a diameter of 31~mm. The diameter of 31~mm was chosen for the compatibility of the STRASSE liquid hydrogen target with the STRASSE silicon tracker, while other dimensions for the target cell can be also manufactured, keeping an equivalent effective target cell volume of up to 125~mL. The target cell is placed in the reaction chamber, along the beamline as shown in Fig.~\ref{fig:STRASSE}. The design of the target holder and the extension arm were chosen to fit inside the STRASSE silicon tracking system and its mechanical structure~\cite{STRASSEpaper}.\par
The target cell is made out of thermo-formed Mylar  of 175~$\mu$m thickness. It is built out of a tubular piece thermo-formed from a flat Mylar sheet and glued along one side, an end cap with rounded edges and an entrance window, see Fig.~\ref{fig:target-cell}. The entrance window has a diameter of 20~mm and has a similar shape as the end cap. The Mylar components are glued together using an epoxy resin (3M DP190) that keeps elasticity at cryogenic temperatures. The glued areas have $\approx$ 5~mm overlap and the glue layer has a thickness of $\approx$ 90~$\mu$m with a standard deviation of 4~$\mu$m. Consequently, the thickness of the cell in the glued areas scales up to 2 respectively 3 layers of Mylar and corresponding 1 respectively 2 layers of glue. The target holder is made out of stainless steel (316).\par

\begin{figure}[ht]
    \centering
    \includegraphics[width=0.8\linewidth]{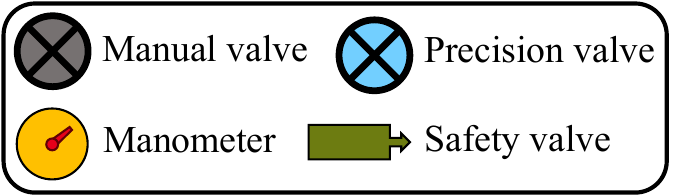} \hfill
    \includegraphics[width=0.85\linewidth]{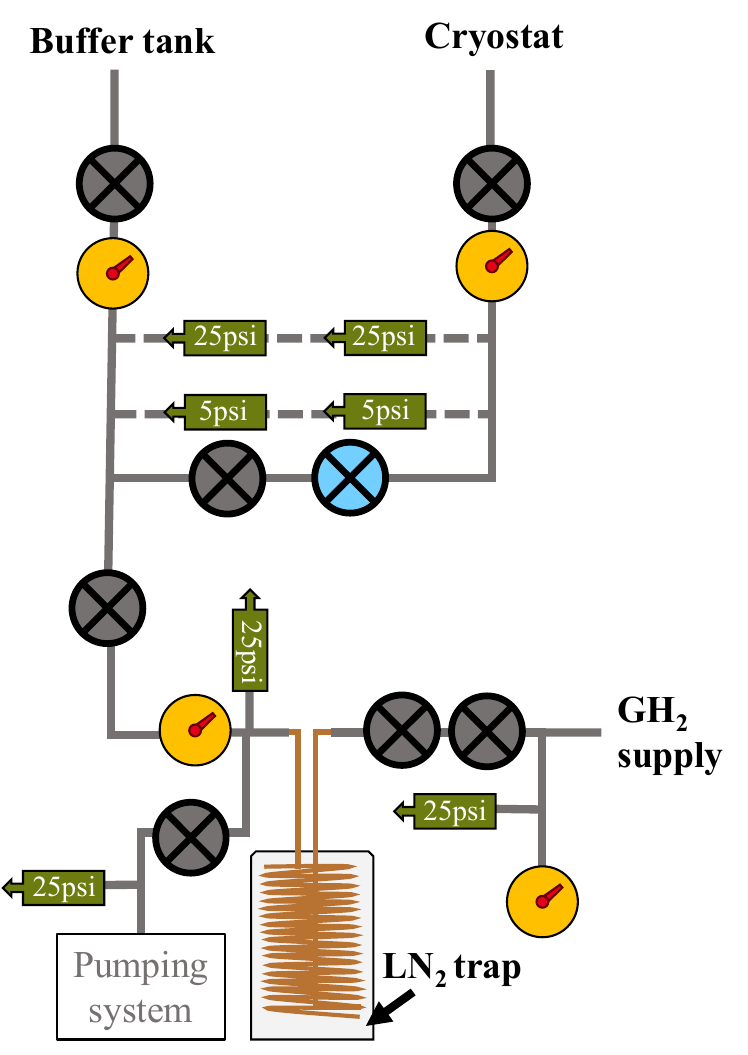}
    \caption{A simplified schematics of the gas-handling system connecting the high-pressure GH$_{2}$ supply, the buffer tank, the cryostat, the pumping system, and the exhaust line. The components of the gas system are labeled and listed in the top legend. The dashed lines indicate the gas route from the cryostat to the buffer tank as a safety measure in case of overpressure in the system. }
    \label{fig:gas-system}
\end{figure}

\subsection{Cryostat}\label{subsec2-2}
\begin{figure*}[t]
    \centering
    \includegraphics[width=1\linewidth]{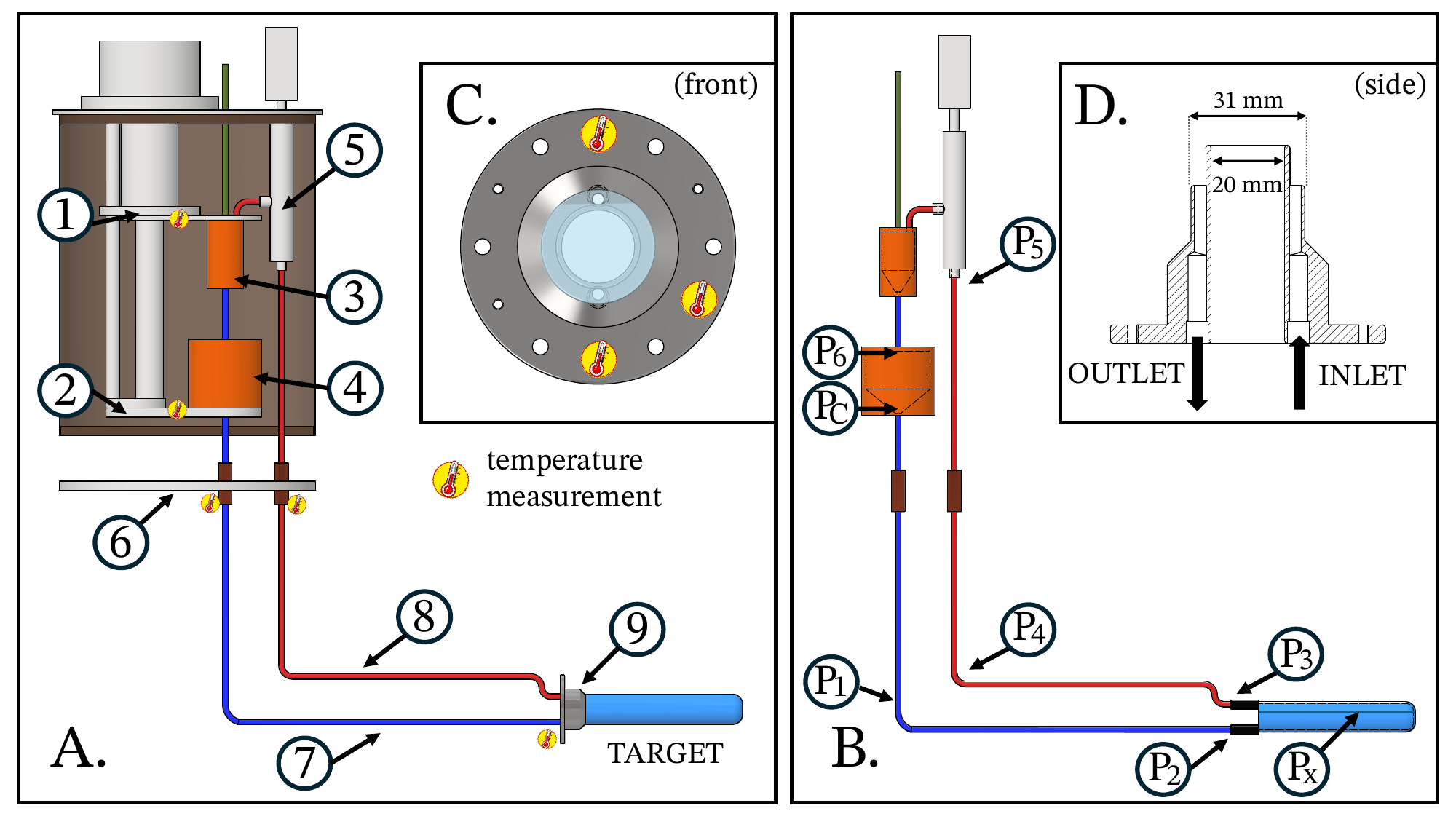}
    \caption{A -- Simplified drawing of the cryogenic system. The image offers an unfolded view of the placement of components inside the cryostat chamber. The components of the cryogenic system labeled on this figure are: the first stage for the cold head (1), the second stage of the cold head (2), the cold trap (3), the condenser (4), the cold valve (5), the feed-through flange between the cryostat chamber and the target chamber (6), the supply line (7), the return line (8), and the target (9). The position of the temperature sensors is indicated by the yellow markers. B -- The thermosiphon loop with labels for the check points used in the thermosiphon model (see Section~\ref{sec4}). For the thermosiphon model calculations, the target holder dimension was not considered, but the heat load due to the target holder was applied in the black regions shown in the sketch. C -- The front view of the target with the indicated position of the three temperature sensors (top, middle, and bottom). From this view angle, the filling level of the target could be monitored during the in-lab tests. D -- the section view of the target holder, with the inlet and outlet pipe holes as well as entrance and total diameter labeled.}
    \label{fig:cryosystem_components}
\end{figure*}

For cooling the hydrogen down to 15-20~K, a pulse tube cold head was used (Cryocooler model PTD406C), manufactured by TransMIT Gesellschaft f\"ur Technologietransfer mbH, Gießen, Germany (Cryo.TransMIT)~\cite{TransMITCryo,Falter2017}. A schematic drawing of the cryogenic system including the cryostat and the target cell is shown in Figs.~\ref{fig:STRASSE} and \ref{fig:cryosystem_components}. The pulse-tube cryocooler has two cooling stages and a rotary valve in remote configuration for minimizing the vibrations in the system~\cite{Thummes1998}. The two stages of the cryocooler can reach temperatures down to 46~K and 6~K, respectively. For the operating temperature range of 50-60~K for the first stage and 15-20~K for the second stage, the cryocooler has a power of up to 10~W at the first stage and 10-15~W at the second stage. The cryostat is operated together with a Sumitomo F-70H helium compressor.\par
A condenser vessel with a volume of $\simeq$220~ml is mounted at the second stage of the cold head. The temperature of the condenser can be monitored and controlled by a silicone diode temperature sensor and a resistive heater. The target cell and the condenser are connected via two copper pipes (ETP), the supply and the return line, with an inner diameter of 4~mm. The supply line and the return line pass from the cryostat chamber to the reaction chamber via a feed-through flange. To reduce the thermal coupling between the supply/return pipes and the feed-though flange, an additional copper tube housing is installed, shown schematically in Fig.~\ref{fig:cryosystem_components}. The design of the cryostat was also optimized to minimize its size and height relative to the beamline. A system of 1~m overall height was obtained.\par
The cryogenic system is additionally equipped with a cold valve shown in Fig.~\ref{fig:cryosystem_components}. An empty target measurement followed by a fast refill of the hydrogen target cell is possible by closing and reopening this cold valve. Details are given in section~\ref{subsec2-2}.\par
\subsection{Gas system}\label{subsec2-3}
The hydrogen gas is supplied to the cryostat by the gas-handling system shown in Fig.~\ref{fig:gas-system}. The gas-handling system is composed of the purifying section where the hydrogen gas is guided through a liquid nitrogen trap when filling the buffer tank from an external gas bottle to freeze out all impurities onto the walls of the copper spindle; the pumping section, allowing to make vacuum in the whole piping system and inside the target cell using a pre-vacuum pump (Edwards nXDS i10); and the \hbox{500-L} buffer tank which is initially filled with 1.5~bar, resulting in a total of 750~L of hydrogen at STP (Standard Temperature and Pressure) in the whole hydrogen circuit. During all operations, the hydrogen is supplied to the cryostat only from the buffer tank and it is returned to the buffer tank after warm-up.\par
\subsection{The working principle}\label{subsec2-4}
The liquid hydrogen target system is based on the thermosiphon loop principle. There are 4 main components in a thermosiphon loop: the \emph{condenser}, the \emph{supply} pipe, the \emph{return} pipe, and our target cell is the \emph{evaporator} component. They are shown in a simplified schematics in Fig.~\ref{fig:cryosystem_components}.\par
The liquid hydrogen drops from the condenser along the supply line into the target cell where the liquid hydrogen gradually warms up followed by evaporation in the region of the target cell. The pressure in the system depends on the setpoint temperature, chosen by the user, at the condenser. The highest-pressure point in the system is the bottom point of the supply line. From the condenser down the supply line, the pressure increases due to the height difference. Along the horizontal section of the supply line, there is no height difference anymore so there is only a pressure drop caused by the viscosity of the fluid and form losses. Moving on from the target cell along the return line, the pressure keeps decreasing by friction, form losses, and height difference. The vapor pressure after the target cell is larger than the vapor pressure in the condenser, which creates a gradient of pressure driving the hydrogen vapors back into the condenser where the cycle restarts. \par
For our purpose, the system needs to be designed in such a way that the vaporization happens above the target cell (i.e. in the return line) and the target cell is full of liquid with a minimum amount of bubbles. This can be controlled by optimizing the height difference between the target cell and the condenser in the design phase of the system, by optimizing the amount of hydrogen used in the system during operation, and by finding the optimum setpoint temperature for the condenser during operation. \par
The target cell, target holder and the supporting arm are thermally insulated from the cold head. The only way in which the target cools down is by means of the cold hydrogen which flows in the closed circuit. Additionally, the target is receiving heat load from the surrounding reaction chamber which is at room temperature and from the silicon tracking system whose electronic boards can reach 45$^{\circ}$C. For a vacuum level of 10$^{-6}$~mbar maintained in the reaction chamber, the heat load from the surrounding detectors and chamber is of radiative nature. Additional heat load comes from the contact of the thermosiphon loop components with the mechanical structure at the feed-through flange and at the target holder. On the target cell side, several layers (5-10) of aluminized Mylar of 6~$\mu$m thickness are used to cover the sides of the target cell, the target holder and the supporting structure of the target holder including the supply and return pipes. The front face of the target cell on the other hand, is not covered by thermal insulation for reducing the material budget on the path of the fragment isotopes after the nuclear reaction. This brings a considerable amount of heat load into the target cell in the form of radiative heat from the surrounding elements such as detectors and the room-temperature reaction chamber. Nonetheless, due to the relatively good level of vacuum in the target chamber, the uncovered front face of the target cell did not show signs of an undesired layer deposition due to acting as a cold trap. Based on theoretical calculations and model tuning (detailed in Section 4.1), the estimated heat load on the uninsulated front face of the target cell is approximately 220~$W/m^2$. In contrast, the side walls of the target cell and the pipes, protected by multiple layers of aluminized Mylar, receive a significantly reduced load of approximately 5~$W/m^2$ and 1~$W/m^2$, respectively.\par
During the development and testing phases, the filling level was directly monitored and confirmed via video recordings of the target cell's front face. These visual observations allowed for determining a pattern of the temperature sensors' reading on the target holder, which serves as the primary indicators of a full target during experimental runs where visual access is restricted.\par
\section{Operation}\label{sec3}
\begin{figure}
    \centering
    \hfill\includegraphics[width=0.88\linewidth]{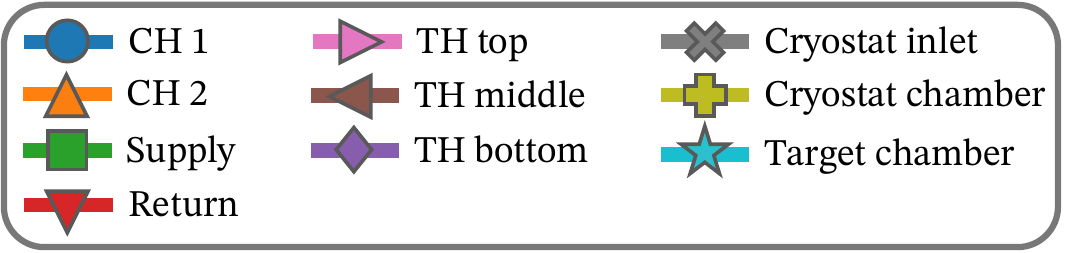}
    \includegraphics[width=\linewidth]{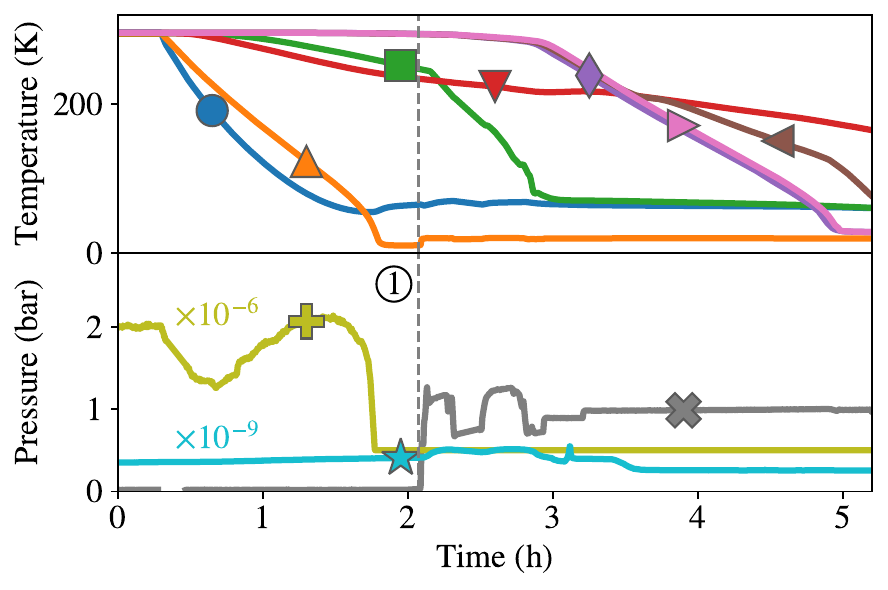}
    \includegraphics[width=\linewidth]{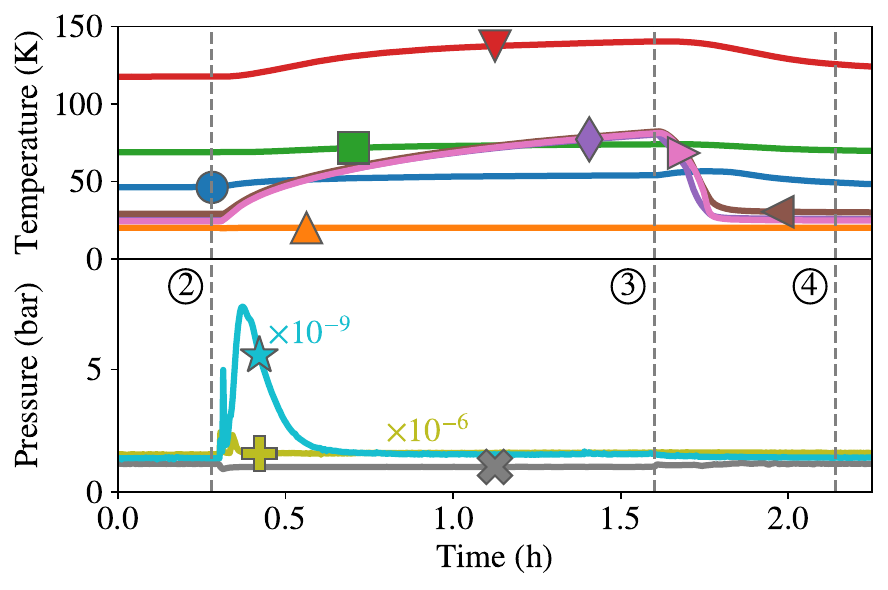}
    \includegraphics[width=\linewidth]{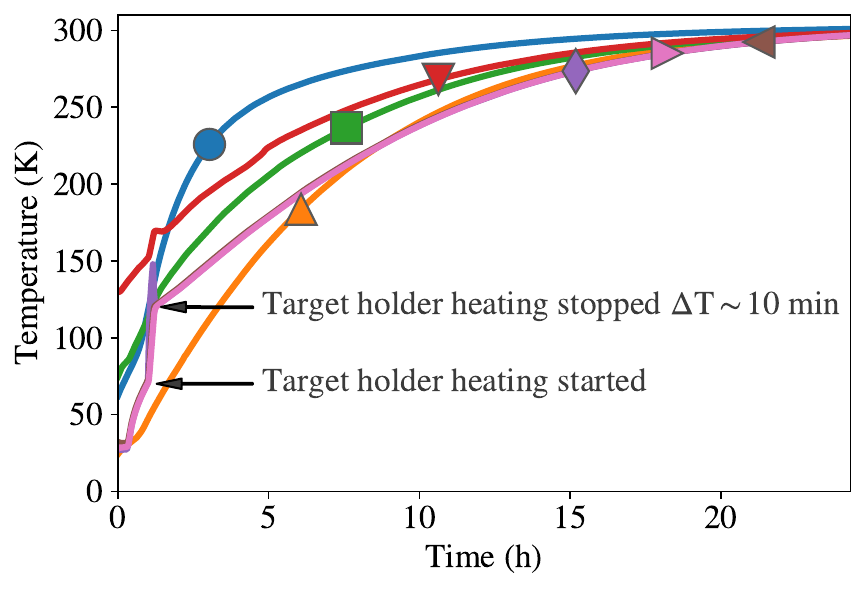}
    \caption{Time evolution of various temperature and pressure values of the cryogenic system during cool-down (top), empty-target mode (middle), and warm-up (bottom). The top legend shows the color code for the following measured quantities: temperature at the first stage of the cold head (CH1), temperature of the second stage of the cold head (CH2), temperature of the supply line -- on the copper insulation housing (Supply), the temperature of the return line -- on the copper insulation housing (Return), the temperature of the top, middle, and bottom positions of the target holder (TH), and the pressure measurements at the cryostat inlet, cryostat chamber and target chamber. The dashed lines mark the following user actions: start of H$_{2}$ filling (1), close the cold valve (2), open the cold valve (3), and the moment when the target cell is full of liquid after an empty-target phase (4).}
    \label{fig:operation}
\end{figure}
\subsection{Monitoring and controlling system}\label{subsec2-1}
\paragraph{Equipment}
The LH$_2$ cryogenic target system is equipped with a series of temperature sensors, in locations shown in Fig.~\ref{fig:cryosystem_components}, as part of the monitoring system. Silicon diode temperature sensors (LakeShore DT-670) and platinum temperature sensors (LakeShore PT111) are used. These are read out by a LakeShore Temperature Monitor 228 module. At the first stage of the cold head, a PT111 temperature sensor and a pair of two \hbox{50-$\Omega$} heaters are installed. These two heaters are only used when a fast final warm-up of the system is needed. At the second stage of the cold head, which is the coldest point in the system, a main DT-670 sensor backed-up by a second redundant sensor and a pair of two \hbox{50-$\Omega$} heaters are used. For maintaining the temperature at the second stage of the cold head at the desired value, the heater output power is controlled based on the temperature sensor's reading in a feedback loop using a LakeShore Monitor 335 module. Moving further down in the system, a PT111 temperature sensor is mounted on the copper housing of each of the supply and return lines. A \hbox{100-$\Omega$} heater is also mounted on the feed-through flange for the warm-up phase. The target holder is equipped with three DT-670 temperature sensors mounted as shown in Fig~\ref{fig:cryosystem_components}-C; one sensor is mounted on top, close to the return line, one sensor is mounted at one third the height of the target (middle sensor), and the third sensor is mounted on the bottom of the target holder. Note that the top and bottom sensors are always mounted closer to the hydrogen pipes while the middle sensor is the furthest from the hydrogen flow. For this reason, the middle temperature sensor measures larger values than the top and bottom ones. Two \hbox{100-$\Omega$} heaters are mounted on the target holder for the warm-up phase. The used heaters are thick-film type of  \hbox{50-$\Omega$} (AP101 50R F) and \hbox{100-$\Omega$} (AP101 100R F) from Ohmite. All mounted heaters are connected in series with a thermostat switch (65$^{\circ}$C) to avoid overheating. The DT-670 sensors have a certified calibration (LakeShore) in the temperature range of 1.4~K to 325~K resulting in an uncertainty up to 10~mK in the 10-20~K temperature region and up to 35~mK uncertainty in the 20-300~K temperature region.\par
The pressures on the cryostat filling line and in the buffer tank, indicated in Fig.~\ref{fig:gas-system}, are also monitored during system operation via Swagelok pressure transducers with a pressure reading range of vacuum to 4~bar. Similarly, the vacuum levels in the cryostat chamber and the reaction/target chamber are readout by Pfeiffer modules (TPR 280 and PKR 361) and logged throughout the full operation of the system. Moreover, the helium compressor of the cryostat is integrated in the monitoring and control system, which allows to turn on and off the cooling remotely in case of emergency. The gas-handling system can only be operated manually.  \par
\paragraph{Safety}
The system is equipped with several safety layers for each emergency scenario. The most concerning scenario is the clogging of the thermosiphon circuit. This could happen due to solidification of hydrogen or other impurities. To avoid the solidification of hydrogen, the second stage of the cold head (i.e. the coolest point of the circuit) is controlled by a pair of heaters and a cryogenic temperature sensor. To avoid the solidification of other impurities, the hydrogen is purified before introducing it into the system and a low vacuum level is ensured before filling the cryostat with hydrogen. If the pressure still increases in the system, a system of check-valves (5~psi and 25~psi, see Figure~\ref{fig:gas-system}) is mounted on the route from the cryostat to the buffer tank to allow the gas to return to the buffer tank and release the excess pressure. The position of the check valves is marked by dashed lines in Fig~\ref{fig:gas-system}. Additionally, the target cell was tested to be able to withstand a pressure of 9~bar at cryogenic conditions or up to 12~bar at room temperature. In case of a rapture of the target cell, the reaction chamber is equipped with a burst disk (1.2 -4~bar) to release the overpressure into the atmosphere. Moreover, all safety exhaust pipes and outlets of pumps are routed to a common exhaust suitable for hydrogen and flammable gases.\par
\subsection{Operation procedures}\label{subsec2-2}
Before the cool-down and the insertion of the hydrogen into the circuit, a good vacuum level (10$^{-3}$~mbar) is required in the gas system, the target cell and cryostat inner piping, buffer tank and the cryostat chamber, while a vacuum level of 10$^{-6}$~mbar is required in the reaction chamber. After the buffer tank is filled with purified hydrogen, the compressor connected to the cryostat can be turned on for begining the cool down. Fig.~\ref{fig:operation} (top) shows the evolution of temperatures and pressures in the system over time during cooling. Up to the vertical dashed line (no. 1), the cryostat cools down with no hydrogen gas supplied. After the dashed line, hydrogen is supplied from the buffer tank to the cryostat via the gas-handling system in several steps. When hydrogen is introduced in the system, one can notice a small rise in the pressure of the reaction chamber before the newly introduced hydrogen is cooled down. This can happen due to hydrogen gas permeation through the Mylar walls of the target cell when the temperature is still large. The cryostat filling line pressure is also fluctuating corresponding to the filling of the hydrogen; the pressure increases initially and then decreases as the hydrogen liquefies in the condenser at the chosen setpoint temperature (setpoint temperature of 16~K in this figure during hydrogen filling). Soon after the hydrogen is introduced into the cryostat, the supply line starts to cool down followed by the target holder. After a total of about 5 hours, when the target holder top and bottom reading difference is reaching its minimum, the target cell starts to fill to 100$\%$ or up to the filling level possible for the setpoint temperature and hydrogen amount chosen by the user.\par
\begin{figure}
    \centering
    \includegraphics[width=1\linewidth]{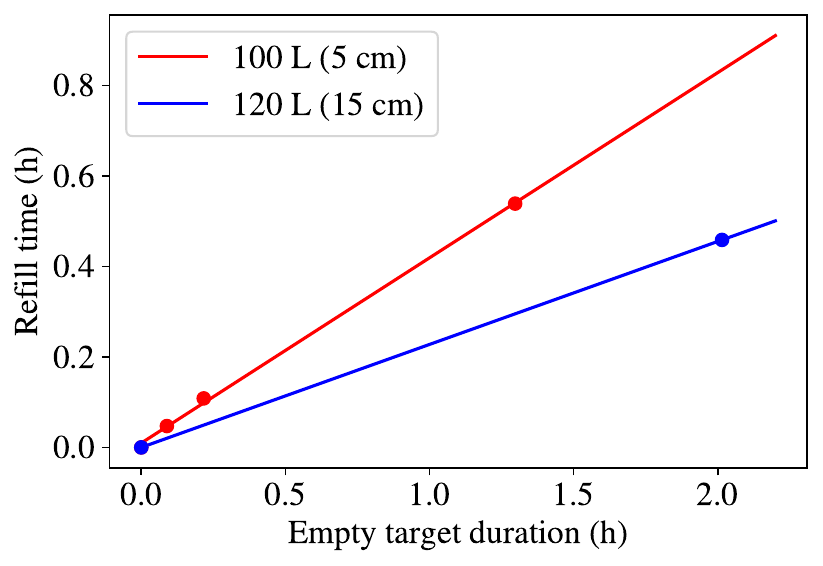}
    \caption{The time needed for a complete target cell refill after an empty-target phase as function of the empty target duration. The data points are given for two different quantities of hydrogen (at STP) in the cryogenic system corresponding to two different target cell lengths: 100~L with 5-cm cell length and 120~L with 15-cm cell length.}
    \label{fig:refill-time}
\end{figure}

The design of the system with the additional cold valve mounted on the return line of the thermosiphon loop allows for empty-target measurements during an experimental beam time. This means that the target cell can be empty (only low density hydrogen vapors present), while the system remains cool, which allows for a fast refill. This operation works by blocking the return line of the thermosiphon loop. Pressure accumulates in the return line and pushes the liquid back into the condenser. The condenser has twice the volume of the target cell in the 15-cm-long configuration. During the empty target mode, the target holder will warm up in the absence of liquid hydrogen. The refill time depends on the empty target mode duration. Fig.~\ref{fig:operation} (middle) shows the temperature and pressure evolution during the empty target mode. The first vertical dashed line (no. 2) marks the closing of the cold valve. After the cold valve is closed, the target cell empties in less than 30~s. One can notice a spike in pressure of the reaction chamber and in the cryostat chamber after the cold valve is closed. A possible explanation for the spike in the reaction chamber is that the permeation through the target cell's Mylar walls of the higher-pressure and relatively warmer hydrogen gas accumulating in the target cell is increasing; a possible reason for the spike in the cryostat chamber comes from the mechanics of the cold valve. The second vertical dashed line (no. 3) marks the reopening of the cold valve, after which the liquid starts dripping again in the target cell and cooling the target holder back to the normal operation temperature. The third vertical dashed line (no. 4) marks the refilling ($\geq$90$\%$) of the target cell. Fig.~\ref{fig:refill-time} shows the refill time as function of the empty target mode duration for two configurations: 15-cm-long target cell with 120~L (at STP) of H$_2$ and 5-cm-long target cell with 100~L (at STP) of H$_2$. A linear trend is observed and, moreover, a shorter refill time when 120~L (at STP) of H$_2$ are present in the system due to a larger cooling inertia and a larger mass flow rate in the system.\par
At the end of the operation, the compressor is simply turned off and the hydrogen is allowed to return to the buffer tank while it warms up. This procedure takes typically 24~h as shown in Fig.~\ref{fig:operation} (bottom). To speed up the process, one can use the heaters installed in the system; their impact on the warm-up can be seen in Fig.~\ref{fig:operation} (bottom). \par
\subsection{Operation parameters}\label{subsec2-3}
The setpoint temperature in the condenser, together with the amount of hydrogen filled in the system, are the two parameters which can be modified by the user during operation to ensure that the target cell is filled for a given target cell dimension and heat load conditions of the experiment. The allowed setpoint temperature range is 16 to 20~K. Notice in Fig.~\ref{fig:cryosystem_components}, the setpoint temperature is applied on the bottom support for the condenser -- it can be concluded that the temperature of the liquid inside of the condenser is slightly higher based on the higher saturation pressure measured on the cryostat line. With this setpoint temperature range, the saturation pressure in the condenser varies from 0.25 to 1.05~bar. \par
\begin{figure}[h]
\centering    
\includegraphics[width=0.8\linewidth]{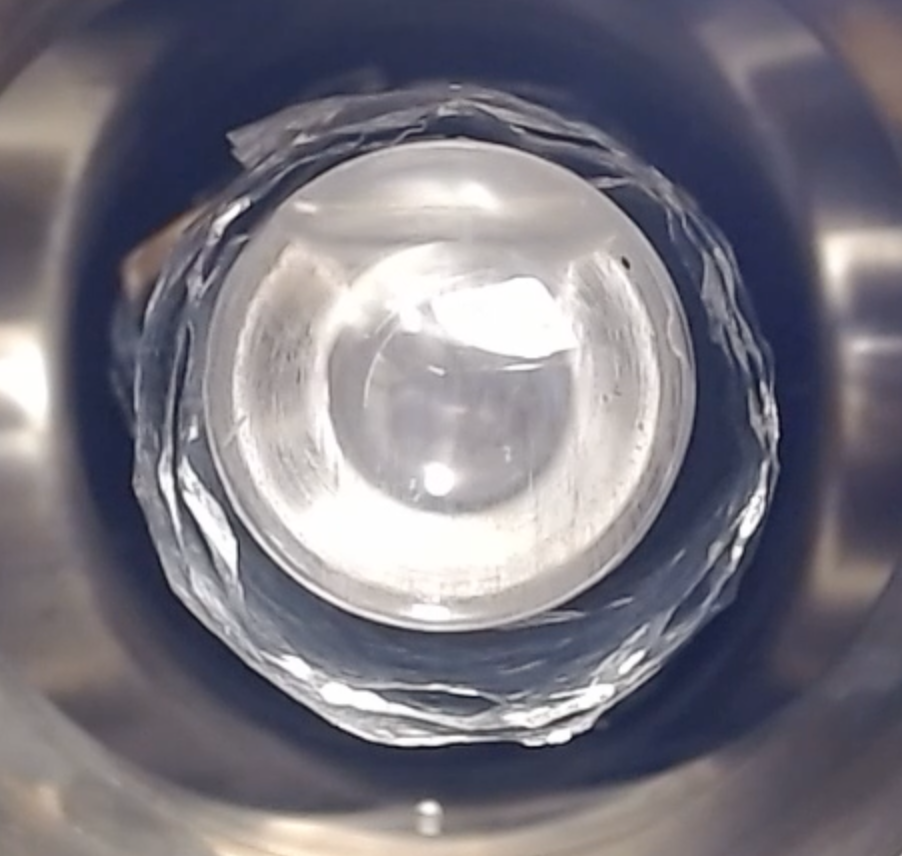}
    \caption{A frame of the recorded data of the front face of the target cell during filling. The level of the liquid is visible in this photo.}
    \label{fig:photo-target-cell}
\end{figure}

The main goal of the user is to have a full target cell of liquid hydrogen. For this purpose, the filling level (FL) as a function of the setpoint temperature and the amount of hydrogen in the system was evaluated. The filling level was extracted based on video recording data for the front face (downstream) of the target cell. A frame from the video recordings is shown in Fig.~\ref{fig:photo-target-cell}. The FL percentage was determined pixel-wise from the selected frames corresponding to each parameter configuration by normalizing the vertical position of the liquid meniscus against the fixed vertical coordinates of the target cell's top and bottom edges: $FL = (y_{liq.}-y_{bot.})/(y_{top}-y_{bot.}) \cdot 100\%$. The width of the vapor-rich boundary layer between the liquid and the gas phase is considered as the filling level uncertainty. \par
\begin{figure}[h]
    \centering
    \includegraphics[width=\linewidth]{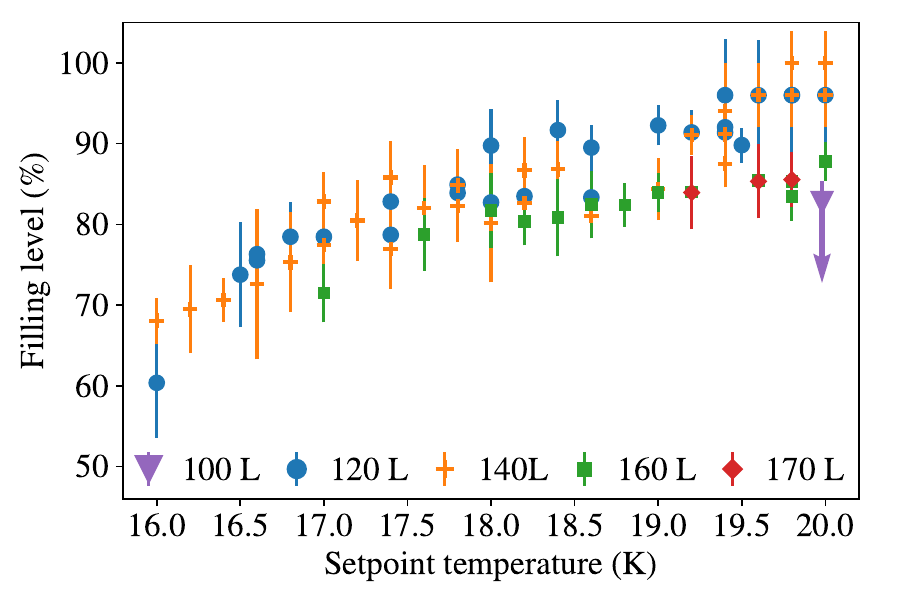}
    \centering
    \includegraphics[width=\linewidth]{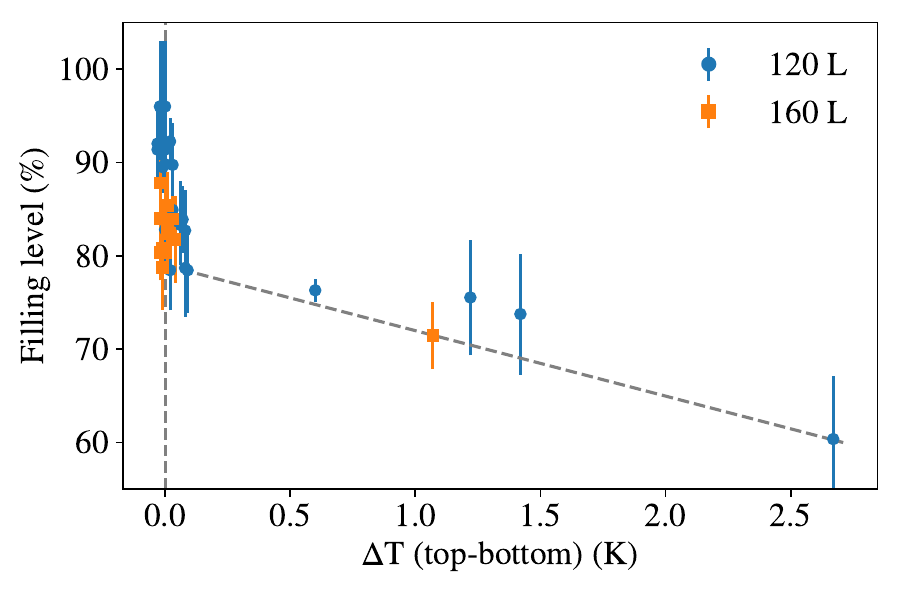}
    \caption{Top: the visible filling level as a function for the setpoint temperature and the amount of hydrogen in the system (different markers, see legend) for a 15-cm long target cell. For 100~L of H$_2$ the full range of setpoint temperatures was not fully scanned, only the point for the highest temperature (20~K) is shown, while for smaller temperature, the filling level would also be smaller, as indicated by the arrow. The error bars indicate the the width of the vapor-rich boundary layer between the liquid and the gas phase. Bottom: the relation between the visible filling level and the top-bottom temperature difference measured on the target holder during operation with a 15-cm long target cell and for two different quantities of hydrogen, 120~L and 160~L. The dashed lines are shown for guiding the eye and indicating the linear trend for filling levels smaller than 75$\%$ and the saturation at $\Delta T(top-bottom) = 0$ above 75$\%$ visible filling level.}
    \label{fig:FL_vs_Tsetpoint}
\end{figure}

The dependence of the filling level on the setpoint temperature and the amount of hydrogen in the system is plotted in Fig.~\ref{fig:FL_vs_Tsetpoint}~(top) for a 15-cm-long target cell. One observes that low setpoint temperature values (16~K) result in a low FL, while high setpoint temperature values (19.5-20~K) result in a full target with \hbox{FL $\geq90\%$}. Moreover, only with 120~L and 140~L of hydrogen (volume at STP), a full target is obtained. A lower amount of hydrogen does not suffice to fill the combined volume of the target cell ($\simeq$ 120~mL), the pipes volume and the bottom of the condenser, see the data point for 100~L. By going to a larger amount of hydrogen a filled target cell is not obtained either, see the data points for 160-170~L in Fig.~\ref{fig:FL_vs_Tsetpoint}~(top). A probable explanation for this is that the cooling power of the cryostat does not suffice to cool down the larger hydrogen mass. \par

Based on the experimental observations, the difference between the top sensor and the bottom sensor of the target holder is a good indication for the filling of the target cell. Fig.~\ref{fig:FL_vs_Tsetpoint}~(bottom) shows the filling level (from video recordings) as a function of the top-bottom temperature difference, $\Delta T_{(top-bottom)}$. Note that one can discriminate between filling levels below 75$\%$. The `saturation' of $\Delta T_{(top-bottom)}$, well before the visible filling level approaches 100$\%$, indicates that the \emph{visible} filling level corresponds to a small vaporization percentage \emph{by mass}, i.e. the vapors above the visible filling level have a high vapor density close to 100$\%$.\par
In the experimental conditions, a visual check of the filling level is not possible, as a result, the $\Delta T_{(top-bottom)}$ can be considered a good indicator for the filling level due to the placement of the top and bottom sensors close to the return and supply lines, respectively.\par

\section{Thermosiphon Model}\label{sec4}
For understanding the observations of Subsection~\ref{subsec2-3} qualitatively, a model was created calculating the filling level in the target cell, the mass flow rate in the loop, as well as the pressure, temperature and vapor quality at several points in the system as a function of the temperature set at the condenser and the hydrogen amount within the cryostat. The heat load was treated as a free parameter in this model, which was tuned to fit the experimental data points. Additionally, the parameter model space was restricted to remain within the cooling power performance of the cold head.\par
The primary motivation for the thermosiphon model is to provide a qualitative and quantitative understanding of the system’s response to varied operating parameters. Specifically, the model clarifies the non-intuitive relationship where increasing the condenser temperature actually improves the filling level by increasing vapor pressure and driving the liquid column further into the cell.\par
\subsection{Physics model}
\paragraph{Fluid properties}
The fluid properties used in the model are functions of temperature and pressure extracted from the \emph{CoolProps} python database~\cite{CoolProp}. A homogeneous model is considered for the fluid found in a mix of liquid and gaseous phase: 
\begin{equation}
\begin{aligned}
\rho_{mix} &= \alpha \rho_{gas} + (1 - \alpha) \rho_{liquid}, \\
\mu_{mix}  &= \alpha \mu_{gas}  + (1 - \alpha) \mu_{liquid},\\
c_{mix}    &= \alpha c_{gas}    + (1 - \alpha) c_{liquid}
\end{aligned}
\label{homog_model_eq}
\end{equation}
\\
where the density ($\rho$), viscosity ($\mu$), and specific heat ($c$) of the gas and that of the liquid are functions of pressure and temperature and $\alpha$ is the vapor quality (by mass) $\alpha = m_{gas}/(m_{gas}+m_{liquid})$. 
The pressure $P$, temperature $T$ and $\alpha$ are calculated in several points along the thermosiphon loop. The starting point is the setpoint temperature in the condenser, chosen by the user, and the pressure in the condenser, considered as the saturation pressure corresponding to the chosen setpoint temperature. The physical state ($P$, $T$, $\alpha$) is calculated at several points along the thermosiphon loop which can be visualized in Fig.~\ref{fig:cryosystem_components}, inset B.
The calculation points are the following:
\begin{itemize}
    \item the bottom of the supply line (point 1)
    \item the end of the supply line (point 2)
    \item the filling level in the target cell (point x)
    \item the exit from the target cell (point 3)
    \item on the bottom of the return line (point 4)
    \item at the top of the return line (point 5)
    \item and back in the condenser (point 6)
\end{itemize}
The volume and shape of the condenser, target cell and the pipes are considered with high fidelity as in the real setup.
\paragraph{Pressure drop} The sum of pressure drop between each consecutive points along the thermosiphon loop must be zero:
\begin{equation}
\Delta P_{C-1}+ \Delta P_{1-2}+ \Delta P_{2-3}+ \Delta P_{3-4}+ \Delta P_{4-5}+ \Delta P_{5-6} = 0.
\label{deltaP_eq}
\end{equation}
The solution for the mass flow rate and the level of liquid in the target cell and in the condenser is found iteratively, to ensure the pressure drop equality shown in Eq.~\ref{deltaP_eq}.\\
The pressure drop between different points in the system can come from hydrostatic pressure difference, friction forces due to the viscosity of the fluid, and form losses due to the 90$^{\circ}$ turn of the pipes or due to crossing from a pipe to a larger volume (target/condenser). The difference in hydrostatic pressure takes the form of
\begin{equation}
\Delta P = \rho g \Delta h,
\end{equation}
where $\rho$ is the density of the fluid, $g$ is the gravitational acceleration, and $\Delta h$ is the height difference between the two points for which the pressure drop is calculate. At each iteration of the calculation, the height of the fluid column in the condenser is recalculated based on the level of fluid in the target cell and the pipes volumes to conserve the total mass of hydrogen.\\
The pressure drop caused by the fluid viscosity along a pipe of diameter $D$ takes the form of
\begin{equation}
\Delta P = f/D\cdot \rho v^2/2,
\end{equation}
with velocity $v = \dot{m}/(\rho A)$ as function of the mass flow rate $\dot{m}$, fluid density $\rho$ and pipe cross-section $A$, respectively; the friction factor $f$ depends on the Reynolds number $Re = \rho \cdot D/\mu$, where $\mu$ is the fluid viscosity. Following the Blasius correlation~\cite{Blasius}, if the Reynolds number is smaller than 2300, then the friction factor equals $f = 64/Re$ for laminar flow, else $f = 0.3164\cdot Re^{-0.25}$ for turbulent flow~\cite{fluidMech}.\\
For the pipes bent at 90$^{\circ}$, the model considers a form loss of 
\begin{equation}
\Delta P = K \cdot \rho v^2/2,
\end{equation}
where $K = 0.3$~\cite{fluidFlow} for the large-curvature turns (curvature larger than twice the pipe diameter) and $K = 0.9$~\cite{fluidFlow} for the sharp corners. The supply line and return line have a smooth curvature in the reaction chamber region while in the cryostat chamber, where the return line connects back into the cold trap and condenser, the pipes undergo sharp turns. Another source of form losses for pressure appears due to the transition from the pipes of diameter of 4~mm to the target cell of diameter of 31~mm or to the condenser of 76~mm. These form losses have the same expression, but with $K = (1-(D_{in}/D_{out})^{2})^{2}$, depending on the inlet and outlet diameter value. 

\paragraph{Heat exchange}
The model considers heat load $q_{pipes}$ along the supply and return line, $q_{TH}$ acting on the inner pipes of the target holder, $q_{target}$ for the side of the target cell covered by thermal insulation and $q_{front}$ for the front side of the target cell which is not covered by thermal insulation. The sources of heat input are mainly of radiative nature onto the target cell. Nevertheless, the full thermosiphon loop system loses cooling power due to the contact with the mechanical structure. Major cooling power losses are observed at the level of the feed-through separating the target chamber and the cryostat chamber through which the supply and the return line are passing. The complexity of the assembly makes it difficult to estimate the net cooling power of the cryocooler and the heat load on the hydrogen fluid. For this reason, the heat exchange parameters $q_{pipes}$, $q_{TH}$, $q_{target}$, and $q_{front}$ were fitted to the observations of filling level as a function of setpoint temperature. The values listed in Table~\ref{q-values} were used in the model.

\begin{table}[h]
    \centering
    \caption{The values for the heat load considered on the pipes, target holder (TH), the side of the target cell, and the front face of the target cell.}
    \label{q-values}
    \begin{tabularx}{\linewidth}{XX}
        \hline
        Parameter & Value \\
        \hline
        $q_{pipes}$  & $1~\mathrm{W/m^2}$ \\
        $q_{TH}$     & $10~\mathrm{W/m^2}$ \\
        $q_{target}$ & $5~\mathrm{W/m^2}$ \\
        $q_{front}$  & $220~\mathrm{W/m^2}$ \\
        \hline
    \end{tabularx}
\end{table}
These values are also explained by a theoretical heat load calculation of black-body (BB) radiation $P_{BB} = \sigma_{Stefan-Boltzmann} \cdot (T_{amb}^4 - T_{cryo}^4)$ which takes values around 440~W$/\mathrm{m}^2$, for $T_{amb} = 25^{\circ}$C and $T_{cryo} = 20$~K. The front face of the target cell is not completely covered by the insulation foil, but roughly half of the solid angle seen by this front side is obstructed by the insulation foil mounted on the side of the target cell and extended downstream. The side of the target cell is completely covered by several layers of insulation foil and so we can approximate that only about 1$\%$ of this heat load reached the fluid from the side of the target, given the reflectivity of at least 90$\%$ of the insulation foil per layer~\cite{martinez2014}. The target holder has a large material budget and is in contact with the supporting arm. Therefore, we expect another major heat load from the target holder. Finally, the pipes are covered with insulation foil but, nonetheless, come in contact with the feed-through flange.\\
Between each pair of points along the thermosiphon loop, the corresponding heat load acts on the fluid, increasing its temperature by $\Delta T = q / (\dot{m}c_{liq.}$), where $c$ is the specific heat of the fluid. If the fluid's temperature becomes larger than the saturation temperature corresponding to the pressure on that segment, vaporization starts. During vaporization, the temperature remains constant and the vapor quality $\alpha$ increases by 
\begin{equation}
\Delta \alpha = q/(\dot{m} \lambda),
\end{equation}
with $\lambda$ being the latent heat value corresponding to the pressure in the current segment ($\lambda$ ranges between 440 and 454 kJ/kg). If the fluid is fully vaporized, $\alpha = 1$, and if the available heat load allows, the fluid in the gaseous form continues to increase in temperature by $\Delta T = q / (\dot{m}c_{vap.}$). This logic applies to all segments of the thermosiphon loop, including the volume of the target cell where the liquid-vapor interface, or the filling level is expected.\par
While the current thermosiphon model utilizes a constant heat load to fit the experimental filling level, it is important to note that the exothermic ortho-para hydrogen conversion contributes an additional, time-dependent internal heat source, not included in the current calculations. At the STRASSE operating temperature of 20~K, the ortho-para conversion heat constant is approximately 527~kJ/kg~\cite{Leachman2017}, a value 20\% larger than the latent heat of vaporization. Assuming an initial `normal' hydrogen composition (75\% ortho) and a total hydrogen amount of 140~L at STP, this conversion contributes approximately 11.8~mW of internal heat at the beginning of the operation, amounting to $\approx 7~\%$ of the radiative heat load received on the front face of the target cell. This heat load decays over time as the para-fraction increases, dropping to approximately 3.4~mW after 100 hours and 1.6~mW after 200 hours of operation. While the spontaneous second-order decay constant is $\approx 0.0114 \text{ hr}^{-1}$~\cite{Milenko1974}, the presence of paramagnetic elements in the stainless steel may further accelerate this process~\cite{Yang2024}. Consequently, the heat load baseline of the system is subject to a continuous drift during the first several days of operation.\par
\begin{figure*}[h]
    \centering
    \includegraphics[width=0.46\linewidth]{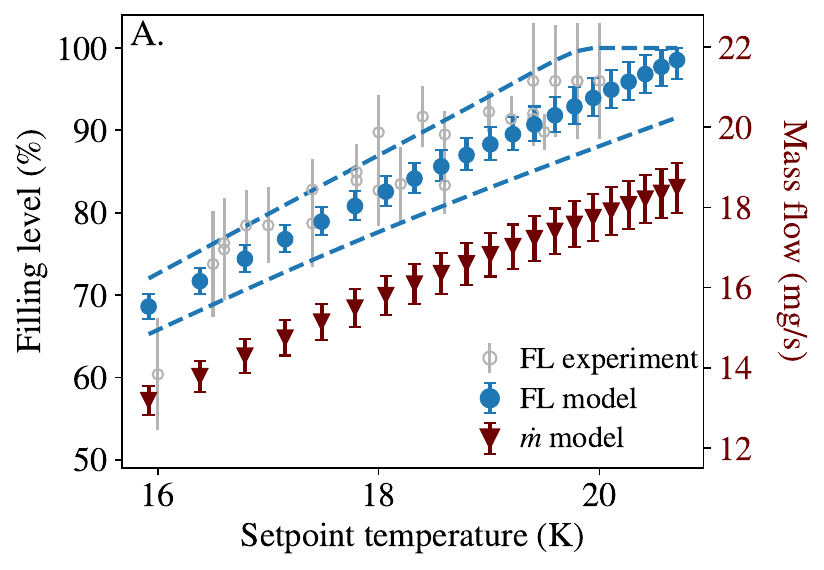}
    \centering
    \includegraphics[width=0.46\linewidth]{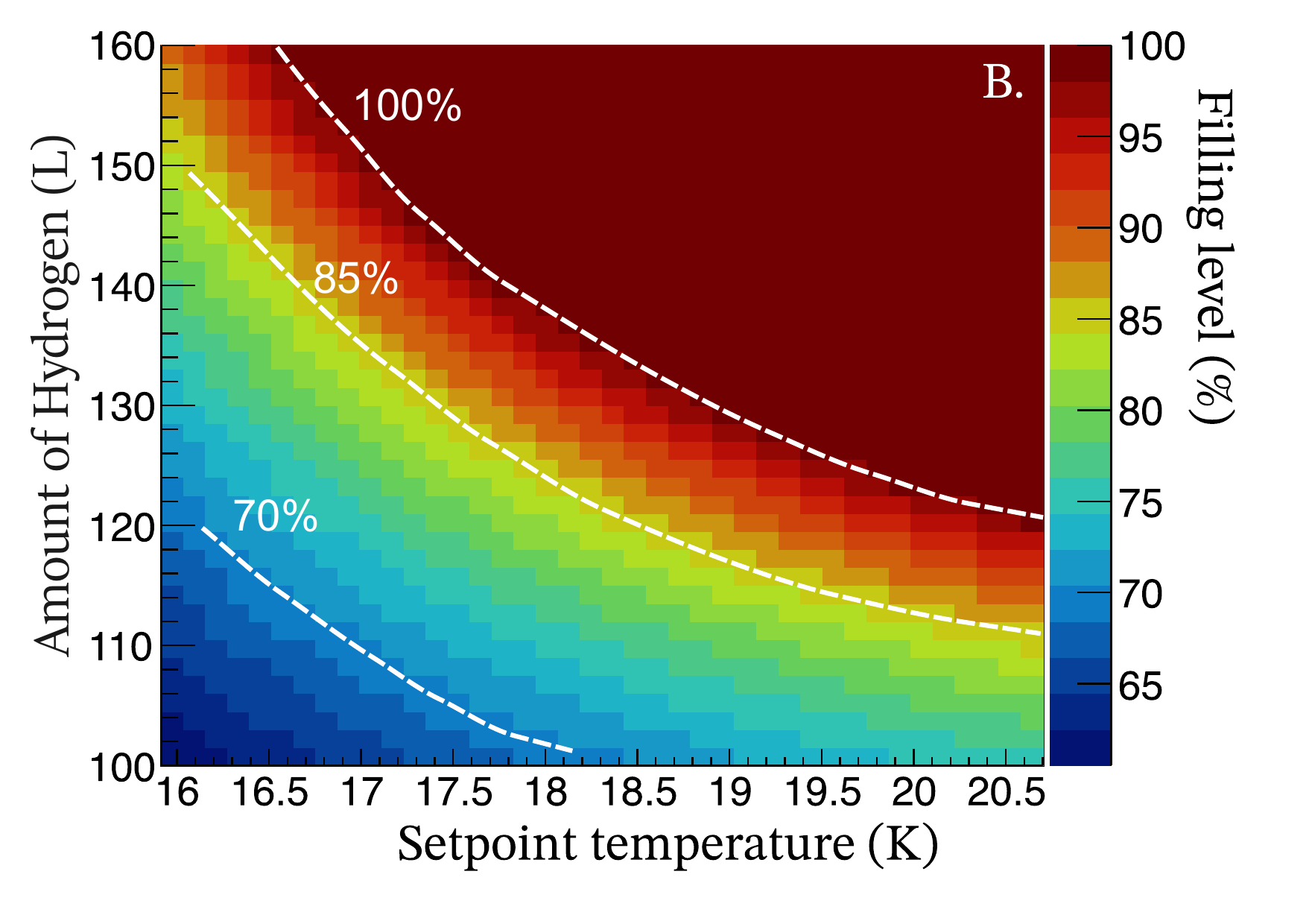}
    \centering
    \includegraphics[width=0.46\linewidth]{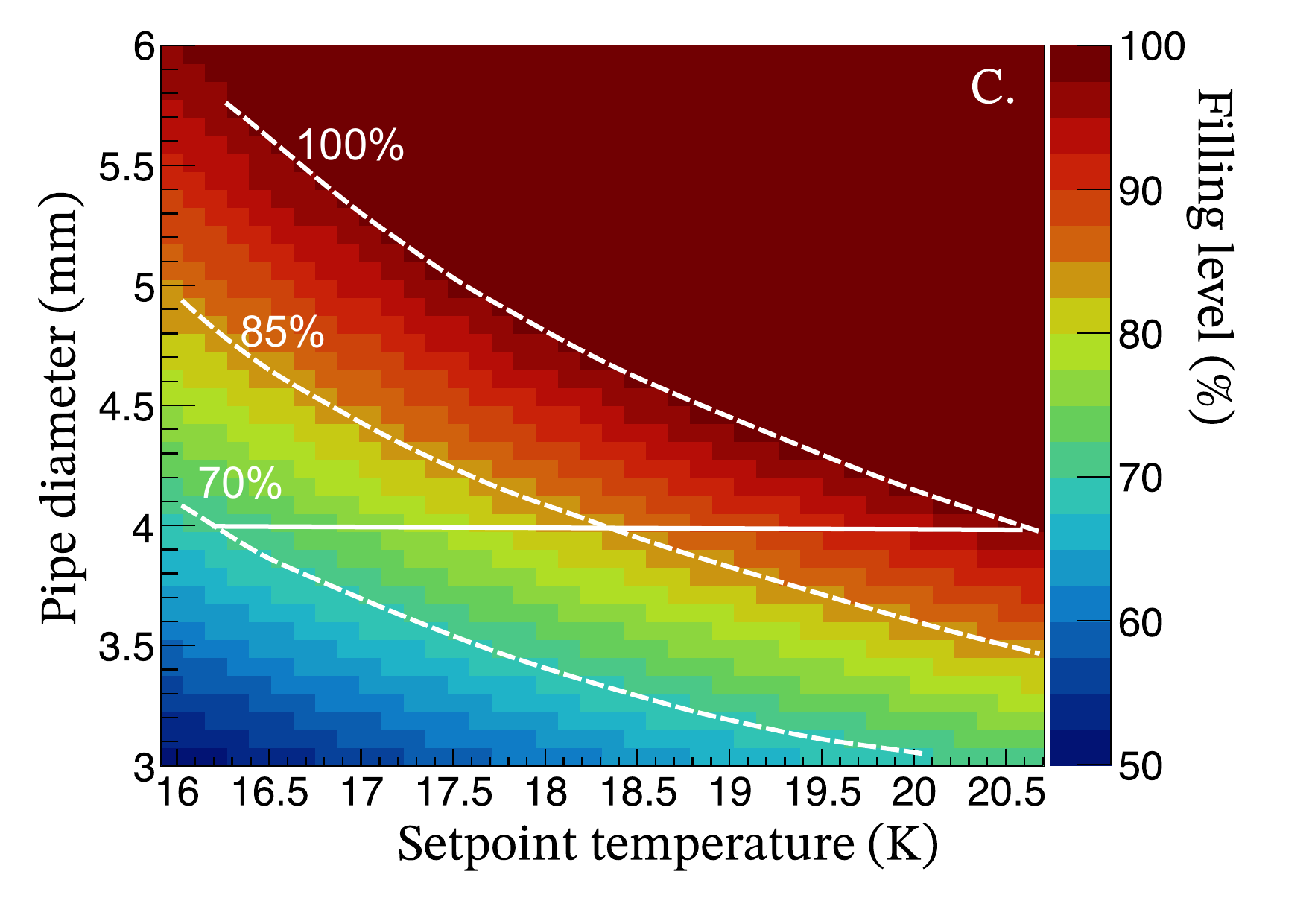}
    \centering
    \includegraphics[width=0.46\linewidth]{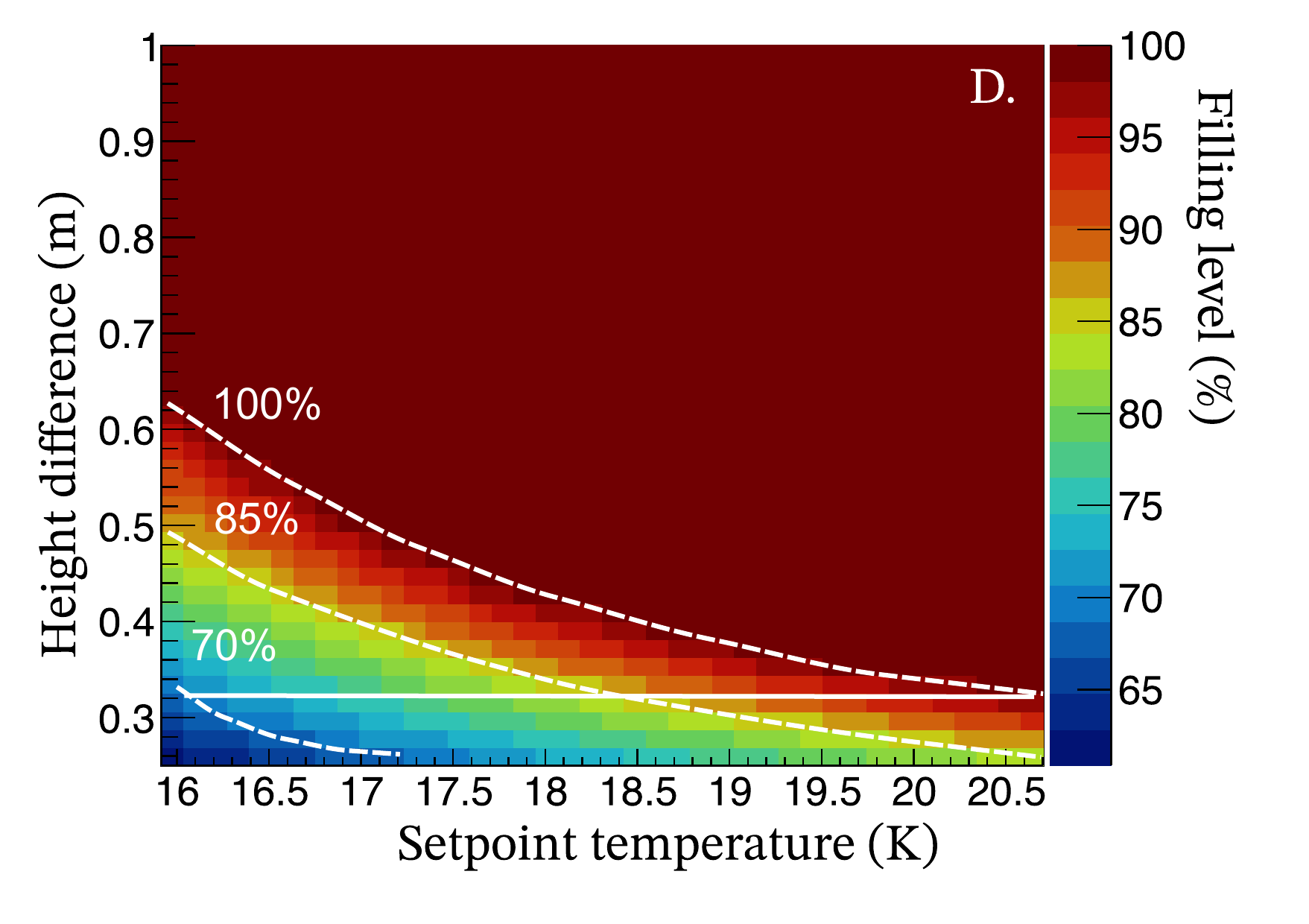}
\caption{A. -- The calculated (model) FL as function of the setpoint temperature compared to the experimental observations (experiment) and the calculated mass flow rate. The error bars represent the relative variation of the heat load by 5$\%$ in the model. The dashed-line contours for the filling level calculations show the uncertainty given by the relative variation of the vaporization percentage threshold by 5$\%$, see text for details. B. to D. -- Calculations for the filling level as function of the setpoint temperature and the amount of hydrogen in the system (B., for pipe diameter = 4~mm and height difference = 0.32~m), the diameter of the supply and return pipes (C. for 120~L of hydrogen and height difference = 0.32~m), and the height difference between the condenser and the target cell (D. for 120~L of hydrogen and pipe diameter = 4~mm). In all plots, the calculated filling level is considered where the vaporization percentage becomes 5$\%$ by mass (75$\%$ vaporization by volume). The horizontal solid lines in plots (C.) and (D.) show the current dimension of the system for the pipe diameter and the height difference, respectively. The white dashed curves are displayed for the filling level values of 100\%, 85\%, and 70\% for guiding the eye and improve readability for black-and-white print.}
    \label{fig:toy-model-fillinglevel-vs-setpoint}
\end{figure*}
\subsection{Implementation}
\paragraph{Algorithm}
The solution for the system, ($\dot{m}$, $FL$, {$P$}, {$T$}, {$\alpha$})$_{sol}$, composed of the mass flow rate, the filling level, and the array of pressure, temperature, and vapor quality values in all calculation points of the thermosiphon loop, is found iteratively. The initial conditions ($\dot{m}$, $FL$, {$P$}, {$T$}, {$\alpha$})$_{IC}$ are set based on the chosen setpoint temperature and its corresponding vapor pressure in the condenser. The filling level is searched at each iteration using the previous $\dot{m}$ value, followed by the search of a new $\dot{m}$ and the calculation of all ($P$, $T$, $\alpha$) values in all check points. The level of liquid in the condenser is adjusted at each iteration to preserve the amount of hydrogen in the system. This loop continues until the solution converges and $\delta\dot{m}/\dot{m}$ and $\delta FL/FL$ reach the tolerance value of $10^{-6}$. The bisection method is used to find the FL and the solution of $\dot{m}$ in each iteration.\par
\paragraph{Parameter tuning}
The calculated liquid-vapor interface is not a sharp boundary, but rather a gradient of vaporization percentage by mass. In order to match the visually observed filling level from our recordings, a threshold for the vaporization percentage needs to be established. We found that our data are best understood with a vaporization percentage threshold of $\alpha_{\text{threshold}} = 5\%$ and the heat load values mentioned in the previous subsection. The heat load and the $\alpha_{\text{threshold}}$ parameters are correlated and not unique for fitting the observed data for filling level vs. setpoint temperature. Nevertheless, they only act as a normalization factor and do not change the slope of the calculated dependence of the filling level on the setpoint temperature.  \\
\subsection{Results} 
The filling level as a function of the setpoint temperature is calculated using the described model and is compared to the experimental observations shown in Fig.~\ref{fig:toy-model-fillinglevel-vs-setpoint} A. One can see that the model describes the trend of the experimentally observed filling level well. The error bars show the variation of the calculated filling level caused by a variation in heat load of 5$\%$ (relative). The dashed lines show the variation in the calculated filling level when varying the vaporization percentage threshold $\alpha_{\text{threshold}}$ by 5$\%$ (relative). Both the heat load and the vaporization percentage threshold are parameters that need to be tuned, but in both cases we can observe that the slope of the filling level as a function of the setpoint temperature does not change. The mass flow rate is also plotted in Fig.~\ref{fig:toy-model-fillinglevel-vs-setpoint} A. As the setpoint temperature is varied from 16~K to 20~K, the the mass flow rate increases from $\sim$13 to 18~mg/s according to the calculations, increasing in the filling level. The given error bars for the calculated mass flow rate correspond to the variation of the heat load by 5$\%$ (relative). \par
For a better understanding of the cryogenic system, one can find in Fig.~\ref{fig:toy-model-fillinglevel-vs-setpoint} B.-D. the filling level for different setpoint temperatures depending on the amount of hydrogen in the system (volume at STP), the diameter of the supply line and the height difference between the condenser and the target cell, respectively.\par
By increasing the amount of hydrogen in the system, one increases the height of the column of hydrogen in the condenser and on the supply line, which in turn increases the hydrostatic pressure at the level of the target cell. A larger pressure difference between the condenser and target cell translates into a higher level of liquid in the target cell. According to the model, one could reach a higher filling level with a larger amount of hydrogen. However, the cooling power of the cryostat is limited and one can reach a point where the large amount of hydrogen can no longer be efficiently cooled down, resulting in a target cell full of bubbles and no longer filled to 100$\%$.
The limitation in cooling power is not included in the model and is only taken into account when considering the allowed heat load onto the system. This behavior was experimentally observed while operating the 15-cm long target, see Fig.~\ref{fig:FL_vs_Tsetpoint}.
For the 15-cm long target cell, the amount of hydrogen in the cryostat must exceed 100~L to obtain a full target cell, also possible with 120~L and 140~L of hydrogen. For even larger quantities, 160-170~L, the filling level does not exceed \SI{90}{\percent}.
At the same time, with a 5-cm long target, one can obtain a 100$\%$ filling level already from 80~L of hydrogen and one has a larger operation range up to 160-170~L where the cooling power limitation can be observed again.\par
The diameter of the supply and return pipes also prove to impact the filling level of the target cell as shown in Fig.~\ref{fig:toy-model-fillinglevel-vs-setpoint} C. According to the model, the mass flow rate doubles in value between the lower left ($\dot{m} = $10~mg/s) and the top right corner ($\dot{m}$ = 22~mg/s) of this figure as the pipe diameter doubles in size. This can be easily understood since the mass flow rate is directly proportional to the pipe diameter for a given fluid velocity and the fluid velocity is constrained by the pressure drop due to form losses in the system. \par
Finally, the height difference between the condenser and the target cell translated directly in a larger pressure gradient and a larger mass flow rate inside the thermosiphon loop. The mass flow rate almost triples in value between the lower left corner and the top right corner of the parameter range of \hbox{Fig.~\ref{fig:toy-model-fillinglevel-vs-setpoint} D.} ($\dot{m}$~=~12 to 32~mg/s as the height difference and setpoint temperature are varied). One can observe that according to our model, a height difference above 0.62~m would mean that the target cell is 100$\%$ full for any setpoint temperature that one uses. \par
Due to the space availability to couple with the STRASSE silicon tracker and for an easy installation in the experimental area, the STRASSE liquid hydrogen target converged to the current design with a height difference of 0.32~m between the target cell and the condenser and 4~mm inner pipe diameter. This compact design fulfills our requirements and allows us to have a full liquid hydrogen target for cell lengths of up to 150~mm. \par
\subsection{Discussion}
The heat dissipation efficiency of a thermosiphon loop is often measured by the temperature difference between the condenser and the evaporator~\cite{Caner2024}. For an efficient thermosiphon loop the heat input onto the fluid is exclusively put into the latent heat and not used for additional temperature increase at the evaporator. For our use case, an efficient thermosiphon loop coincides with a regime with 100$\%$ filling level, as the main source of heat input is on the target cell. If the target cell would only be partially filled with liquid, it would mean that the hydrogen gas above the liquid level would receive heat load and increase its temperature beyond the boiling temperature. The installed temperature sensors on our setup do not allow for a direct measurement of the fluid's temperature in the condenser or in the target cell. Nevertheless, even in this situation, the temperature difference between the target holder and the second stage of the cold head was observed to be at a minimum when using 140~L in an operation with a 15-cm long target cell (80-100~L for a 5-cm target cell) and at the setpoint temperature of 20~K, or higher (a higher setpoint temperature is in principle possible, but the pressure regime would exceed 1~bar).\par
\begin{table}[h]
    \centering
    \caption{Mean temperature and pressure values together with the corresponding standard deviation for the presented in-beam use over 2.5~days in Fig.~\ref{fig:stability}.}
    \label{tab:stability}
    \begin{tabularx}{\linewidth}{X r | X r}
        \toprule
        \textbf{Temperature} & & \textbf{Mean (Variation)} & \\
        \midrule
        \multicolumn{2}{l}{CH1} & \multicolumn{2}{r}{46.46(6) K} \\
        \multicolumn{2}{l}{CH2} & \multicolumn{2}{r}{20.19(2) K} \\
        \multicolumn{2}{l}{Supply} & \multicolumn{2}{r}{68.65(9) K} \\
        \multicolumn{2}{l}{Return} & \multicolumn{2}{r}{118.04(57) K} \\
        \multicolumn{2}{l}{TH top} & \multicolumn{2}{r}{24.40(3) K} \\
        \multicolumn{2}{l}{TH middle} & \multicolumn{2}{r}{29.19(10) K} \\
        \multicolumn{2}{l}{TH bottom} & \multicolumn{2}{r}{25.39(4) K} \\
        \midrule
        \textbf{Pressure} & & \textbf{Mean (Variation)} & \\
        \midrule
        \multicolumn{2}{l}{Cryostat inlet} & \multicolumn{2}{r}{1.14(1) bar} \\
        \multicolumn{2}{l}{Cryostat chamber} & \multicolumn{2}{r}{$1.62(3) \times 10^{-3}$ mbar} \\
        \multicolumn{2}{l}{Target chamber} & \multicolumn{2}{r}{$1.26(5) \times 10^{-6}$ mbar} \\
        \bottomrule
    \end{tabularx}
\end{table}
\begin{figure}[h]
    \centering
    \hfill\includegraphics[width=0.88\linewidth]{Figure5and11_legend.pdf}
    \includegraphics[width=\linewidth]{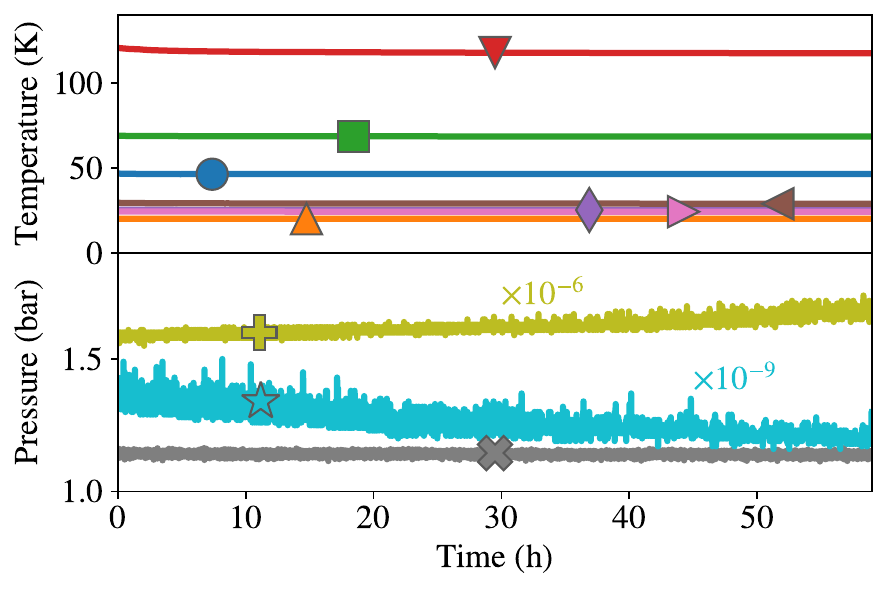}
    \caption{This plot shows the stability of the system (temperature and pressure readings) over 2.5~days of beam time. The shown time sequence comes after a empty target run and the warm-up, which explains the shift in the vacuum level. The mean and standard deviation of the plotted temperature and pressure values are listed in Table~\ref{tab:stability}.}
    \label{fig:stability}
\end{figure}
\section{In-beam validation and performance}\label{sec5}
The STRASSE LH$_2$ target system was validated in-beam in a physics experiment at the RIKEN Nishina Center, Japan. The experiment used the STRASSE LH$_2$ target system with a \SI{5}{cm}-long target cell which was operated for a total of 8 days with 5 days of in-beam use.
The target cell was surrounded by the TOGAXSI silicon array \cite{Tanaka2023} and its electronics. The TOGAXSI silicon array consisted of two silicon wavers each with a size of 78.4 mm (768 strips) by 51.0 mm (494 strips).
Each silicon waver is connected to one APV25-S1 chip. The chip has a power consumption of 2.31 mW per channel which adds up to a total power consumption of $\sim$\SI{2.92}{\watt} per chip. 
The \SI{10}{\litre} STRASSE target chamber and \SI{216}{\litre} TOGAXSI chamber were merged and evacuated to a few \SI{E-6}{mbar}. 
The beam intensity reached up to 4.1$\times$10$^{5}$~cps at F3 focal plane in BigRIPS. The power dissipated by the beam inside the target cell is lower than 1~$\mu$W. For comparison, the radiative heat load on the uninsulated front face of the target is approximately 170 mW (220 $W/m^2$). Since the beam-induced heat is roughly five orders of magnitude smaller than the environmental heat load, local boiling caused by the beam intensity is negligible for our current experimental program.\par
Fig.~\ref{fig:stability} shows the performance of the STRASSE LH$_2$ target system during a steady-state operation of 59 hours, after an empty target run and before the warm-up.
Temperature and pressure values for this steady-state period are listed in Table \ref{tab:stability}.
All temperatures remain stable during in-beam use.
The first and second stage of the cryostat operate within the expected temperature range. The supply and return line temperatures have an offset to the hydrogen temperature in the pipes caused by the positioning of the sensors onto the copper tube surrounding the respective pipe.
The temperature reading of the target holder top and bottom fluctuates with less than \SI{0.04}{K} despite the target cell being surrounded by heat radiated from TOGAXSI electronics. This highlights the importance of the high vacuum in the target chamber. The vacuum of the target chamber decreases by \SI{0.2e-9}{bar} over time, caused by the continued pumping during operation and the target cell freezing impurities onto the Mylar surface.
The cryostat inlet pressure remains constant with a fluctuation of \SI{0.01}{bar}, underlining the stable state of the liquid hydrogen target.\par

\section{Conclusions and perspectives}\label{sec6}
A new liquid hydrogen target was developed for proton-induced knockout reaction experiments at the RIBF facility of RIKEN, Nishina Center, Japan.\par
The LH$_2$ target cell can be extended up to 125~mL in volume and it is made out of 170~$\mu$m-thick Mylar, minimizing energy loss and angular straggling in the target. The cryogenic system is based on a two-stage pulse tube (PTD406C) and it is operated between 16 and 20~K. A fast cooldown of the system and hydrogen liquefaction are reached within the first 5 hours of operation and the system features an `empty-target' mode for in-beam background measurements with a quick refill and recovery of the normal operation in less than 1 hour. \par
A thermosiphon model was developed for the hydrogen circuit based on a thermosiphon loop for a better understanding of the system. The model was in agreement with the observations after normalizing parameter tuning. A key feature of the present setup which the model is able to reproduce is the increase of the filling level as the temperature in the condenser is increased. Additionally, an important conclusion for further optimizations of the thermosyphon loop is that a larger pipe diameter and/or a larger height difference between the condenser and the target cell would ensure a full target cell for the full range of condenser temperatures between 16~K and 20.5~K.  \par
The STRASSE liquid hydrogen target system for proton-induced nuclear reactions experiments offers competitive capabilities and features; the target holder and target cell being able to be exchanged according to each beamtime requirement making it a versatile and modular system. The new target system was commissioned in-lab as well as in-beam for a first physics experiment. The cryogenic system, as part of the STRASSE device will continue to be used for studying the neutron rich nuclear shell structure through direct knockout reactions such as (p,2p) or (p,3p) in inverse kinematics and additionally, it can be coupled with other tracking systems such as TOGAXSI for cluster knock-out reactions at the RIBF facility.\\

\section*{Acknowledgments}
The authors acknowledge the support from the Deutsche Forschungsgemeinschaft (DFG, German Research Foundation) -- Project-ID. 279384907 -- SFB 1245. We are grateful for the valuable support from Jens Falter and Jack Schmidt from Cryo.TransMIT during the cryogenic system development and commissioning. Additionally, we want to thank Daniel K\"orper and Haik Simon from R$^{3}$B, GSI, who hosted our in-lab tests of the cryogenic system and Akane Sakaue for hosting the in-beam commissioning at the RIBF, RIKEN.

\bibliographystyle{unsrtnat}
\bibliography{references}

@article{QFSreview,
   title={Quasifree (p,2p) and (p,3p) reactions with unstable nuclei},
   volume={88},
   ISSN={1089-490X},
   DOI={10.1103/physrevc.88.064610},
   number={6},
   journal={Physical Review C},
   publisher={American Physical Society (APS)},
   author={Aumann, T. and Bertulani, C. A. and Ryckebusch, J.},
   year={2013}}

@article{AO_reviewlh2,
   title={Hydrogen targets for exotic-nuclei studies developed over the past 10 years},
   volume={47},
   ISSN={1434-601X},
   DOI={10.1140/epja/i2011-11105-5},
   number={9},
   journal={The European Physical Journal A},
   publisher={Springer Science and Business Media LLC},
   author={Obertelli, A. and Uesaka, T.},
   year={2011}}

@misc{COCOTIERref,
  title = {COCOTIER project},
  author = {Corsi, Anna},
  howpublished = {IRFU, France},
  url = {https://anr.fr/Project-ANR-17-CE31-0005},
  year = {2017}
}

@article{PRESPECpaper,
title = {The {PRESPEC} liquid-hydrogen target for in-beam gamma spectroscopy of exotic nuclei at {GSI}},
journal = {Nuclear Instruments and Methods in Physics Research Section A: Accelerators, Spectrometers, Detectors and Associated Equipment},
volume = {736},
pages = {81-87},
year = {2014},
issn = {0168-9002},
doi = {https://doi.org/10.1016/j.nima.2013.10.035},
author = {C. Louchart and J.M. Gheller and Ph. Chesny and G. Authelet and J.Y. Rousse and A. Obertelli and P. Boutachkov and S. Pietri and F. Ameil and L. Audirac and A. Corsi and Z. Dombradi and J. Gerl and A. Gillibert and W. Korten and C. Mailleret and E. Merchan and C. Nociforo and N. Pietralla and D. Ralet and M. Reese and V. Stepanov}}

@article{MINOSpaper,
  author    = {Obertelli, A. and Delbart, A. and Anvar, S. and Audirac, L. and Authelet, G. and Baba, H. and Bruyneel, B. and Calvet, D. and Château, F. and Corsi, A. and Doornenbal, P. and Gheller, J.-M. and Giganon, A. and Lahonde-Hamdoun, C. and Leboeuf, D. and Loiseau, D. and Mohamed, A. and Mols, J.-Ph. and Otsu, H. and Péron, C. and Peyaud, A. and Pollacco, E. C. and Prono, G. and Rousse, J.-Y. and Santamaria, C. and Uesaka, T.},
  title     = {{MINOS: A vertex tracker coupled to a thick liquid-hydrogen target for in-beam spectroscopy of exotic nuclei}},
  journal   = {The European Physical Journal A},
  year      = {2014},
  volume    = {50},
  number    = {1},
  pages     = {8},
  month     = {jan},
  doi       = {10.1140/epja/i2014-14008-y}}

@article{STRASSEpaper,
  author    = {Liu, H. N. and Flavigny, F. and Baba, H. and Boehmer, M. and Bonnes, U. and Borshchov, V. and Doornenbal, P. and Ebina, N. and Enciu, M. and Frotscher, A. and Gernh{\"a}user, R. and Girard-Alcindor, V. and Goupilli{\`e}re, D. and Heuser, J. and Kapell, R. and Kondo, Y. and Lee, H. and Lehnert, J. and Matsui, T. and Matta, A. and Nakamura, T. and Obertelli, A. and Pohl, T. and Protsenko, M. and Sasano, M. and Satou, Y. and Schmidt, C. J. and Sch{\"u}nemann, K. and Simons, C. and Sun, Y. L. and Tanaka, J. and Togano, Y. and Tomai, T. and Tymchuk, I. and Uesaka, T. and Visinka, R. and Wang, H. and Wienholtz, F.},
  title     = {{STRASSE}: a silicon tracker for quasi-free scattering measurements at the {RIBF}},
  journal   = {The European Physical Journal A},
  year      = {2023},
  volume    = {59},
  number    = {6},
  pages     = {121},
  doi       = {10.1140/epja/s10050-023-01018-3}}

@article{Tanaka2023,
title = {Designing {TOGAXSI}: {Telescope for inverse-kinematics cluster-knockout reactions}},
journal = {Nuclear Instruments and Methods in Physics Research Section B: Beam Interactions with Materials and Atoms},
volume = {542},
pages = {4-6},
year = {2023},
issn = {0168-583X},
author = {J. Tanaka and R. Tsuji and K. Higuchi and H. Baba and M. Böhmer and T. Furuno and R. Gernhäuser and Y. Hijikata and S. Ishimoto and T. Kawabata and S. Kawase and Y. Kubota and S. Kurosawa and S. Takeshige and T. Uesaka and K. Yahiro and J. Zenihiro},
doi = {https://doi.org/10.1016/j.nimb.2023.05.054}}

@online{TransMITCryo,
  author = {{TransMIT Center for Adaptive Cryotechnology and Sensors}},
  title = {Cryogenic Systems},
  year = {2025},
  url = {https://cryo.transmit.de/en/},
  note = {Accessed: 2025-11-02}
}

@inproceedings{Falter2017,
  author = {Falter, J. and others},
  title = {{Pulse Tube Cryocooler at 4 K: Customization for Sensitive Cryoelectronic Applications in ``Dry'' Low Noise Cryostats}},
  booktitle = {KRYO 2016},
  address = {Freyburg, Germany},
  year = {2016},
  url = {https://snf.ieeecsc.org/files/ieeecsc/2023-05/revedhdrFalter%2C%20Jens_TransMIT-Poster.pdf}
}

@article{Thummes1998,
title = {{Small scale 4He liquefaction using a two-stage 4K pulse tube cooler}},
journal = {Cryogenics},
volume = {38},
number = {3},
pages = {337-342},
year = {1998},
issn = {0011-2275},
author = {G Thummes and C Wang and C Heiden},
doi = {https://doi.org/10.1016/S0011-2275(97)00169-0}}

@article{CoolProp,
  title = {{Pure and Pseudo-pure Fluid Thermophysical Property Library CoolProp}},
  author = {Bell, Ian H. and Wronski, Jorrit and Quoilin, Sylvain and Lemort, Vincent},
  journal = {Industrial \& Engineering Chemistry Research},
  volume = {53},
  pages = {2498--2508},
  year = {2014},
  doi = {10.1021/ie4033999}
}

@Inbook{Blasius,
author="Blasius, H.",
title="Das Aehnlichkeitsgesetz bei Reibungsvorg{\"a}ngen in Fl{\"u}ssigkeiten",
bookTitle="Mitteilungen {\"u}ber Forschungsarbeiten auf dem Gebiete des Ingenieurwesens: insbesondere aus den Laboratorien der technischen Hochschulen",
year="1913",
publisher="Springer Berlin Heidelberg",
address="Berlin, Heidelberg",
pages="1--41",
isbn="978-3-662-02239-9",
doi={10.1007/978-3-662-02239-9\_1}}

@book{fluidMech,
  title={Fluid Mechanics},
  author={White, F.},
  isbn={9781259165924},
  series={McGraw-Hill series in mechanical engineering},
  year={2015},
  publisher={McGraw-Hill Education}
}

@manual{fluidFlow,
  title        = {Flow of Fluids Through Valves, Fittings, and Pipe},
  author       = {{Crane Co.}},
  organization = {Crane Co.},
  address      = {Stamford, CT},
  year         = {1999},
  note         = {Technical Paper No. 410}
}

@article{Caner2024,
title = {Simulation of a two-phase loop thermosyphon using a new interface-resolved phase change model},
journal = {International Journal of Heat and Mass Transfer},
volume = {228},
pages = {125607},
year = {2024},
issn = {0017-9310},
author = {Julien Caner and Etienne Videcoq and Adel M. Benselama and Manuel Girault},
doi = {https://doi.org/10.1016/j.ijheatmasstransfer.2024.125607}}

@book{Leachman2017,
    title={Thermodynamic Properties of Cryogenic Fluids},
    author={Leachman, Jacob W. and Jacobsen, Richard T. and Penoncello, Steven G. and Lemmon, Eric W.},
    isbn={978-3-319-57835-4},
    year={2017},
    publisher="Springer Cham",
    doi = {10.1007/978-3-319-57835-4}
}

@article{Milenko1974,
  author  = {Yu. Ya. Milenko and R. M. Sibileva},
  title   = {Natural ortho-para conversion in liquid hydrogen},
  journal = {Ukrainskii Fizicheskii Zhurnal},
  volume  = {19},
  pages   = {2008},
  year    = {1974},
}

@article{Yang2024,
  author = {Yang, Liujing and Li, Xinbao and Chen, Ying and Zheng, Xiaoling and Sun, Kai},
  title = {Ortho to para hydrogen conversion over bimetallic iron and cobalt catalysts},
  journal = {Scientific Reports},
  volume = {14},
  number = {1},
  pages = {20925},
  year = {2024},
  month = {Sep},
  day = {09},
doi = {10.1038/s41598-024-71790-9},
issn = {2045-2322}}

@article{martinez2014,
  title={Experimental study of multilayer insulation {(MLI)} for cryogenic applications},
  author={Martinez, R. and Fesmire, J. E. and Augustynowicz, S. D.},
  journal={Cryogenics},
  volume={64},
  pages={151--157},
  year={2014},
  doi={10.1016/j.cryogenics.2014.02.006}}

\end{document}